\documentclass[usenatbib]{mnras}
\usepackage{epsfig}
\usepackage{amsmath,amssymb}
\usepackage{color}
\usepackage{booktabs}
\usepackage{pdflscape}

\usepackage{epsfig}
\usepackage{color}
\usepackage{aas_macros}  
\usepackage{amsmath,amssymb}
\usepackage{chngcntr}
\graphicspath{{figures/}}

\newcommand{\Msolar}{\mbox{\,$\rm M_{\odot}$}}        
\newcommand{\Rsolar}{\mbox{\,$\rm R_{\odot}$}}        
\newcommand{\Lsolar}{\mbox{\,$\rm L_{\odot}$}}        

  \newcommand{\Teff}{\mbox{\,\em T$_{\rm eff}$}}         
  \newcommand{\sg}{\mbox{\,log $g$}}                     
 \newcommand{\teff}{\mbox{\,$T_{\rm eff}$}}      

  \newcommand{\kmsec}{\,\mbox{$\mbox{km}\,\mbox{s}^{-1}$}}    
  \def\simge{\mathrel{\raise1.16pt\hbox{$>$}\kern-7.0pt
    \lower3.06pt\hbox{{$\scriptstyle \sim$}}}}           
  \def\simle{\mathrel{\raise1.16pt\hbox{$<$}\kern-7.0pt
    \lower3.06pt\hbox{{$\scriptstyle \sim$}}}}           
\newcommand{\Zstar}{\mbox{\,EC\,19529$-$4430}}
\newcommand{\Jstar}{\mbox{\,GALEX\,J184559.8$-$413827}}

\title[EC 19529--4430 - a metal-poor EHe star]{EC 19529--4430: SALT identifies the most carbon- and metal-poor extreme helium star}

\author[C. S.~Jeffery et al.]{C. S. Jeffery$^{1}$\thanks{email: simon.jeffery@armagh.ac.uk},
L. J. A. Scott$^{1}$,
A. Philip Monai$^{1,2}$,
B. Miszalski$^{3}$ and
V. M. Woolf$^{4}$
\\
$^{1}$Armagh Observatory and Planetarium, College Hill, Armagh, BT61 9DG, UK\\
$^{2}$School of Mathematics and Physics, Queen's University Belfast, Belfast, BT7 1NN, UK\\
$^{3}$Australian Astronomical Optics - Macquarie, Faculty of Science and Engineering, Macquarie University, North Ryde, NSW 2113, Australia\\
$^{4}$Physics Department, University of Nebraska at Omaha, 6001 Dodge St, Omaha, NE, 68182, USA
}

\begin{document}

\date{Accepted \ldots. Received \ldots; in original form \ldots}

\pagerange{\pageref{firstpage}--\pageref{lastpage}} \pubyear{2014}

\maketitle

\label{firstpage}

\begin{abstract}
\Zstar\ was identified as a helium-rich star in the Edinburgh-Cape Survey of faint-blue objects and subsequently resolved as a metal-poor extreme helium (EHe) star in the SALT survey of chemically-peculiar hot subdwarfs. 
This paper presents a fine analysis of the SALT high-resolution spectrum.  
\Zstar\ has $T_{\rm eff} = 20\,700 \pm250$\,K, $\log g /{\rm cm\,s^{-2}} = 3.49\pm0.03$, and an overall metallicity some 1.3 dex below solar; surface hydrogen is $\approx 0.5\%$ by number. 
The surface CNO ratio 1:100:8 implies that the surface consists principally of CNO-processed helium and makes \Zstar\ the coolest known  carbon-poor and nitrogen-rich EHe star. 
Metal-rich analogues include V652\,Her and \Jstar.
Kinematically, its retrograde orbit indicates membership of the galactic halo.
No pulsations were detected in TESS photometry and there is no evidence for a binary companion. 
\Zstar\ most likely formed from the merging of two helium white dwarfs, which themselves formed as a binary system some 11 Gyr ago. 
\end{abstract}


\begin{figure*}
 \includegraphics[width=\textwidth,trim={0cm 0cm 0cm 0cm},clip]{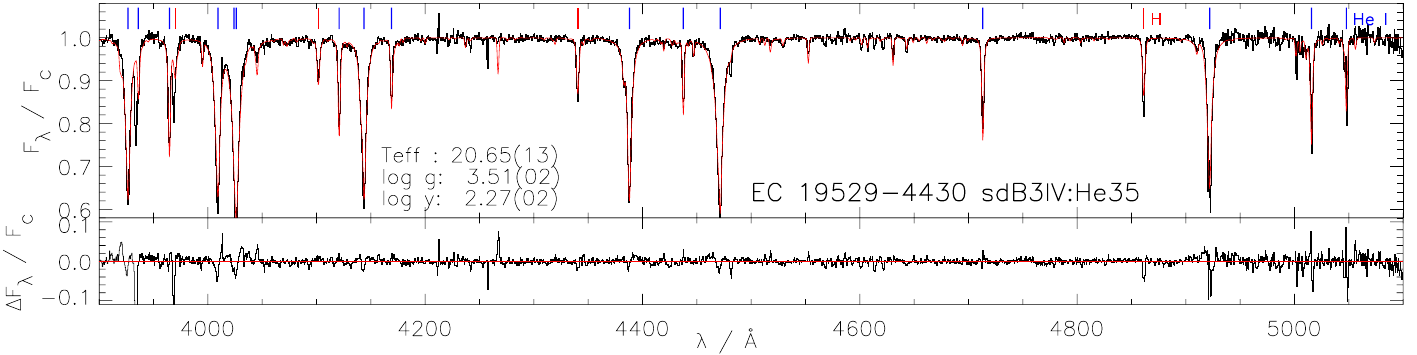}
        \caption{The RSS spectrum of \Zstar\ (black) with the best-fit model (red)  having $T_{\rm eff} = 20\,650 \pm 130$\,K,  $\log g /{\rm cm\,s^{-2}} = 3.51\pm 0.02$,  and $n_{\rm H} = 0.0053\pm0.0002$ ($\log y = 2.27\pm0.02$) (formal errors). The lower panel shows the residuals.
        Hydrogen Balmer and neutral helium absorption lines are marked in red and blue, respectively. }
        \label{f:rss}
\end{figure*}

\section{Introduction}
\label{s:intro}

Extreme helium stars (EHes) are characterised by their spectral types A and B,  weakness or absence of hydrogen Balmer lines, and strength of neutral helium lines. 
Being rare and luminous, their lifetimes must be short, and yet they are located in all major components of the galaxy \citep{philipmonai24}.
High carbon and/or nitrogen abundances indicate evolved surfaces revealed through a catastrophic event in their history, generally thought to be the merger of two white dwarfs \citep{saio00,saio02}. 
Their surface composition and distribution suggest a connection with other hydrogen-deficient stars including the cooler R\,Coronae Borealis variables (RCB) \citep{jeffery94.ccp7a}, dustless hydrogen-deficient carbon stars (dlHdC) \citep{warner67,crawford23,tisserand24a} and the hotter hydrogen-deficient hot subdwarfs (Hesd) \citep{saio00,zhang12a} and O(He) stars \citep{reindl14}.  

Following a spectroscopic survey of luminous blue stars \citep{drilling80,drilling83}, the number of confirmed EHes remained at roughly 17 from the early 1980s until the commencement of the Southern African Large Telescope (SALT) survey of chemically-peculiar hot subdwarfs in 2016 \citep{jeffery21a} (SALT1 hereafter). 
The Edinburgh-Cape (EC) survey of faint blue stars \citep{Cat.EC1,Cat.EC2,Cat.EC3,Cat.EC4,Cat.EC5} classified a number of stars as He-sdB, a classification which overlaps the sdOD stars identified by \citet{green86} and the EHe stars as discussed by \citet{drilling13}.  
SALT1 observations of He-sdB stars from the EC and other surveys revealed 3 new low-gravity B-type EHe stars, namely \Jstar (=J1846$-$4138), \Zstar, and EC\,20236--5703 \citep{jeffery17b,jeffery21a}. 
Concurrently, the \citet{tisserand22} survey of dustless hydrogen-deficient carbon stars identified 2 new A-type EHes, namely A208 (J182447.94--221429.1) and A798 (J183357.03+052917.0).
Meanwhile, several helium-rich subdwarfs were recognised to have properties similar to the hot high-gravity EHe star LS\,IV$+6^{\circ}2$ \citep{jeffery98c} including  PG\,1415+492 \citep{ahmad03a} and PG\,0135+243 \citep{moehler90a}.  
Stars such as BPS\,CS\,22940--0009 \citep{snowdon22}, EC\,20187--4939 \citep{scott23}, and others \citep{naslim10}, suggest that a continuous sequence extends to the helium-rich sdO stars. 

A preliminary analysis of  \Zstar\ ($\alpha_{2000} = 19{\rm h}\,56{\rm m}\,30.7{\rm s}$, $\delta_{2000} = -44^{\circ}\,22\arcmin\,19\arcsec$,
showed it to be the coolest star in SALT1. 
At low-resolution, the spectrum ``shows no ionized helium lines (including He{\sc ii} 4686 \AA) and Balmer lines much weaker than the neutral helium lines. A defining feature is the weakness of all metal lines and the narrow wings of the H and He I lines'' \citep{jeffery21a}. 
\citet{philipmonai24} found \Zstar\ to be a halo object in a retrograde orbit. 
The weak metal-lined spectrum and low surface gravity demands a more careful analysis with model atmospheres of appropriate composition and regard to possible departures from LTE. 
The likely age and retrograde orbit deserve further discussion in terms of origin and evolution.  

The observations on which this analysis is based are described in \S\,2. 
The model atmospheres underlying the analysis are described in \S\,3. 
The spectral analysis itself is presented in \S\,4. 
\S\,5 discusses the distance and kinematics. 
\S\,6 discusses the overall properties and evolutionary status and concludes the paper.


\begin{table*}
\caption[Radial Velocity Measurements]
   {Radial velocities measured from blue HRS spectra of \Zstar. ``RSS'' shows the measurement reported in SALT1. Radial velocities for individual observations are measured in two wavelength ranges relative to an average spectrum. ``Mean'' shows the averages and standard deviations. ``Template'' shows the velocities of the average spectrum relative to a model, and the bottom line indicates the adopted barycentric mean velocity. } 
\label{t:vels}
\small
\begin{center}
\begin{tabular}{lllccc}
\hline
Date & Run & JD & $\sigma_{4540}$ & $\delta V$(4500-4700\AA) & $\delta V$(4300-4500\AA) \\
 & & & & \kmsec & \kmsec \\ 
\hline
20180516 & RSS   &                &   & \multicolumn{2}{c}{$5\pm3$ (SALT1) } \\[1mm]
20190323 & H0049 & 2458566.631215 & 0.010 & $ 0.290\pm0.035$ & $-0.531\pm0.301$ \\
20190329 & H0021 & 2458572.619016 & 0.009 & $-0.117\pm0.032$ & $-0.234\pm0.315$ \\
20190329 & H0022 & 2458572.631412 & 0.011 & $-0.151\pm0.047$ & $-0.200\pm0.353$ \\  
20210801 & H0015 & 2459428.269282 & 0.030 & $-1.863\pm0.115$ & $-0.320\pm0.116$ \\
20210823 & H0025 & 2459450.464167 & 0.058 & $ 0.889\pm0.057$ & $ 0.467\pm0.112$ \\
20210825 & H0018 & 2459452.458264 & 0.060 & $ 0.696\pm0.099$ & $ 2.120\pm0.196$ \\
20210829 & H0017 & 2459456.446146 & 0.051 & $-0.913\pm0.254$ & $-0.601\pm0.172$ \\
         &  Mean  &               &       & $-0.17\pm0.96$ & $ 0.10\pm0.96$ \\
         & Template &             & 0.006 & $+9.73\pm0.57$ & $+11.38\pm1.02$ \\
         &          &             &       & \multicolumn{2}{c}{$10.32\pm 0.70$} \\
\hline
\end{tabular}
\end{center}
\end{table*}

\section[]{Observations}
\label{s:obs}

Observations were obtained in 2019 March and April with SALT using both the  High Resolution Spectrograph (HRS) and the Robert Stobie Spectrograph (RSS). 
These are described in Section 2 and Appendix A of SALT1. 
The methods used for data reduction are described in SALT1 (RSS) and by 
\citet{snowdon22} (HRS). 
Additional short exposure HRS observations were obtained during 2021 August and September in order to establish radial-velocity behaviour.   
For these and for all the red-arm HRS observations we have used the SALT pipeline reductions as described by \citet{kniazev16.hrs-rv}. 

The RSS spectrum and a model fit is shown in Fig.\,E.1 of SALT1.  
Fig.~\ref{f:rss} shows the same spectrum together with a revised model solution which will be discussed below.  
The He\,{\sc i} absorption lines are readily apparent and everywhere narrower than in helium-rich hot subdwarfs,  but less sharp than in the lowest-gravity extreme helium stars. 
The Balmer lines are stronger than in most EHe stars, but similar in strength to those seen in J1846$-$4138  \citep{jeffery17b}.
There are no detectable He{\sc ii} lines in the RSS spectrum. 
The spectrum bears similarities to EHe star V652\,Her \citep{jeffery01b,jeffery15b}, including in particular the extremely rich spectrum of singly-ionized nitrogen (Fig.\,\ref{f:rss}).

The wavelength range covered by the HRS spectrum is 3860 -- 5519 \AA\,  and has a S/N ratio in the range 140 -- 200 at a resolution of $\approx 37,000$. 
The orders were rectified, mapped onto a common logarithmically spaced wavelength grid, and merged.  
The observed spectrum was further renormalised using appropriate models to guide the location of the local continuum in regions where strong lines overlap and to facilitate the measurement of abundances from narrow and weak lines. 

Equivalent widths were measured for a substantial list of lines anticipated in early-type stellar spectra. 
For each predicted line, a region of spectrum is displayed and nearby continuum regions are identified manually. 
The upper and lower limits of the line itself are then identified, and the area of the line under the continuum is integrated and converted to an equivalent width ($W_{\lambda}$)
The error is estimated from the variance of data in the local continuum, as described by \citet{snowdon22}. 
Measurements used in subsequent analyses are shown in Appendix\,\ref{s:lines}.  

Radial velocities of \Zstar\ were measured from the HRS spectra as follows and are reported in Table\,\ref{t:vels}.
The signal-to-noise ratio of each spectrum is represented by $\sigma_{4540}$, the variance around the mean continuum between 4531 and 4548\AA. 
A mean spectrum was formed from the sum of the first three HRS observations. 
Velocities were measured by cross-correlation relative to this mean spectrum in two wavelength ranges, namely 4300--4500\AA\ which includes several broad He{\sc i} lines and 4500--4700\AA\ which contains only sharp lines. Barycentric velocity corrections were applied prior to analysis.
The heliocentric radial velocity of the mean spectrum was measured relative to a laboratory template (``Template'' in Table\,\ref{t:vels}).

With a standard deviation $<1$\kmsec\ over 7 observations on timescales of hours, days, weeks and years, there is no evidence for variability that would indicate membership of a close binary or of surface pulsations (``Mean'' in Table\,\ref{t:vels}). 
The mean HRS velocity is close enough to the RSS velocity to be unremarkable. 

Photometric observations were obtained by the {\it TESS} spacecraft during cycles 13 and 27. These show no periodic variation at any frequency between 0.2 and 50 d$^{-1}$ with an  amplitude  $>0.02\%$ (4$\sigma$) of the mean flux. Pulsations are not expected for low-metallicity extreme helium stars with the properties of \Zstar, where sufficient iron is necessary to drive pulsations \citep{jeffery99a}.

\section{Model Atmospheres and Spectrum Synthesis}
\label{s:physics}

Analyses of astronomical spectra require the use of theoretical models adjusted to match the observations. 
In the present case we consider the spectrum to be formed in the atmosphere of a star which is in hydrostatic and radiative equilibrium. 
Other physics assumptions are implicit in the choice of computer code used to construct the models; for these we refer below to papers associated with the codes adopted.  

Most of our models are computed with the codes {\sc sterne} and {\sc spectrum} described by \citet{behara06a} and \citet{jeffery01b}. These assume local thermodynamic equilibrium (LTE) for the 
atomic level populations, but depart from strict LTE\footnote{Strict LTE assumes that the local source function is given by the Planck function.} in that they include electron scattering in the radiative transfer calculation for the source function.  
Bound-free opacities are included using data from the Opacity and Iron Projects \citep{seaton94,hummer93}. 
Line opacities due to 559316 lines \citep{kurucz95} are also included. 
Pressure broadening in the H, He{\sc i} and He{\sc ii} lines is treated using the tables of \citet{vidal73}, \citet{beauchamp97} and \citet{schoning89}, respectively. At low densities, the \citet{beauchamp97} tables are supplemented by Voigt profiles using Stark broadening coefficients from \citet{dimitrijevic84,dimitrijevic90}.
\citet{behara06a} and \citet{kupfer17} compared models computed with {\sc sterne} and {\sc atlas9} and  {\sc atlas12} \citep{kurucz93.cd13,kurucz96}  respectively. They found differences of up to 2--3\% in the line forming region, likely due to differences in the opacities and other model assumptions. 

Analyses of the EHe stars V652\,Her \citep{przybilla05,pandey17}  and BD$+10^{\circ}2179$ \citep{pandey11,kupfer17} showed departures from LTE to be small except for very strong lines. 
In both analyses of V652\,Her, the non-LTE hydrogen measurement was improved by almost eliminating line-to-line discrepancies, whilst the surface gravity measured from He{\sc i} lines was lower by $\sim0.5$ dex than found previously \citep{hill81,lynasgray84,jeffery99b,jeffery01b}. 
In contrast, the surface gravity measured by \citet{kupfer17} for BD$+10^{\circ}2179$ increased by $\sim0.3$ dex relative to previous LTE and non-LTE analyses by \citet{heber83,pandey06a,pandey11}. This is likely due to the adoption of an improved treatment of metal-line blanketing in the {\sc atlas12} model atmosphere and an improved model for the He{\sc i} atom in the line profile treatment.

On the basis of these previous studies, we have assumed for \Zstar\ that departures from LTE will also be small except for very strong lines.
However, since the strong lines affect the measurement of major parameters, our analysis included the effect of departures from LTE on the profiles of hydrogen and helium lines.
First, we computed {\sc sterne} equivalent models in LTE using version 208 of the code {\sc tlusty} \citep{hubeny95,hubeny17a,hubeny17b,hubeny17c,hubeny21}. 
These {\sc tlusty} (LTE) models are consistent with the {\sc sterne} (LTE) models, as established by comparing the temperature structures of the converged models (see \S\,\ref{s:nlte}). 

Using the structure of the previously converged LTE model, we used {\sc tlusty} to solve the statistical equilibrium equations for H, He{\sc i} and He{\sc ii}  without computing any additional temperature corrections. Thus, we computed departure coefficients for the H, He{\sc i} and He{\sc ii} ions with model atoms using 9, 24 and 20 energy levels respectively.
These model ions, sourced from the {\sc tlusty} website,\footnote{http://tlusty.oca.eu/Tlusty2002/tlusty-frames-data.html} used oscillator strengths and photoionisation cross sections from the Opacity Project database \citep{cunto93}, and level energies from NIST \citep{nist20}.
Hence, using the formal solution code {\sc synspec} version 54 \citep{hubeny95,hubeny17a,hubeny17b,hubeny17c,hubeny21}, we computed grids of emergent spectra with these ions in non-LTE ({\sc synspec} non-LTE).
This approach mirrors that used by, for example, \citet{przybilla05} in their analysis of the extreme helium star V652\,Her. 

\begin{figure}
\centering
  \includegraphics[width=.48\textwidth,angle=0,trim={0cm 0.4cm 8cm 0.7cm},clip]{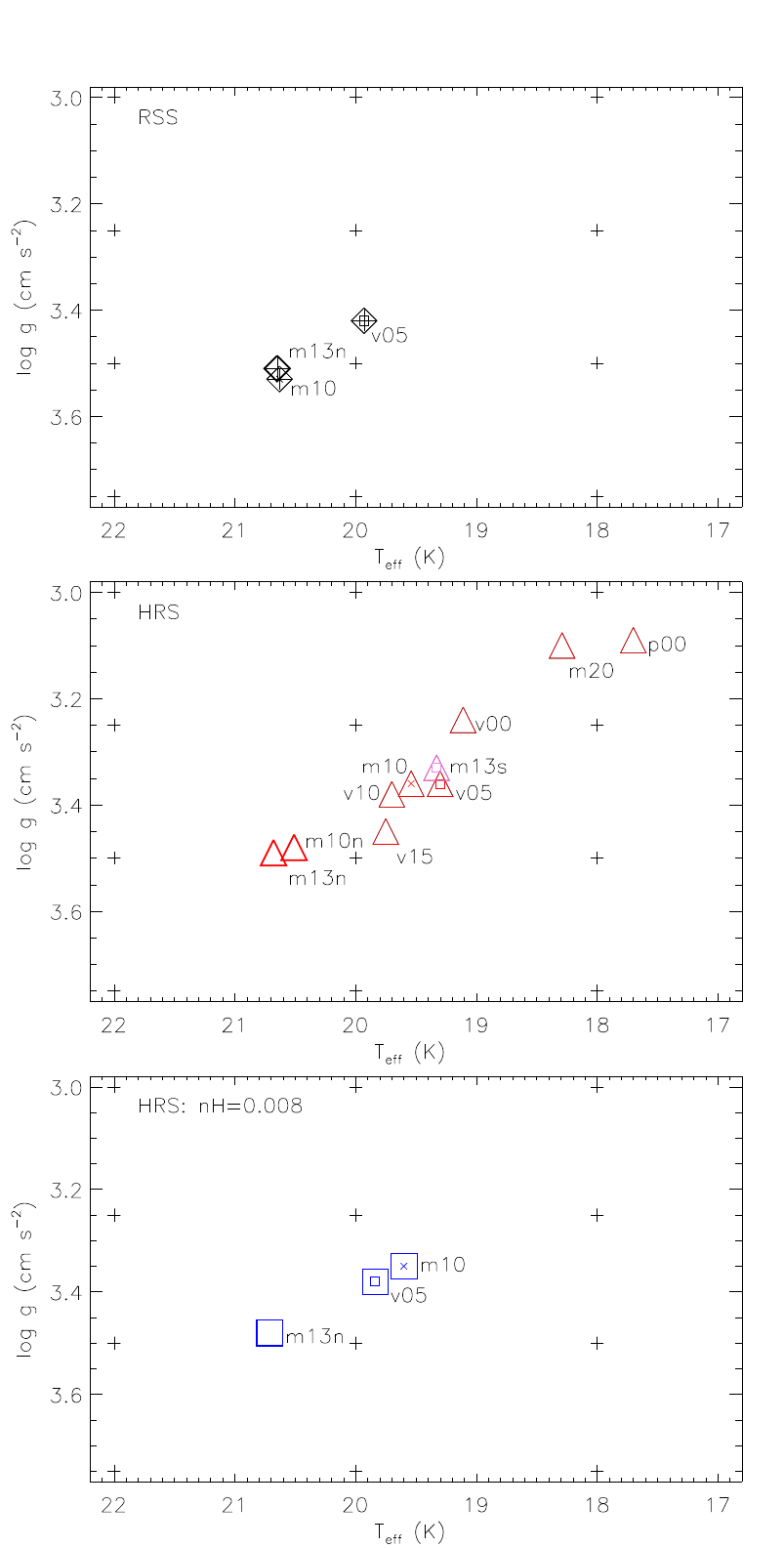}
        \caption{Solutions for $T_{\rm eff}$ and $\log g$ for \Zstar\ for both RSS (top) and HRS spectra (middle). 
        The bottom panel shows solutions for the HRS spectrum with $n_{\rm H}=0.008$.
        Labels {\bf m20}, {\bf m10} (`$\times$') and {\bf p00} indicate fits using the second iteration {\sc sterne} (LTE) $+$ {\sc spectrum} (LTE) grid, with $\xi_{\rm turb}=0$\kmsec.
        Labels v00, v05 (`$\square$'), v10 and v15 indicate fits using the third iteration {\sc sterne} (LTE) $+$ {\sc spectrum} (LTE) grid ({\bf m13n10}), with $\xi_{\rm turb}=0$, 5, 15 and 15 \kmsec.
        Heavy symbols indicate fits using the {\sc tlusty} (LTE) $+$ {\sc synspec} (non-LTE) grids m10n and m13n.
        The violet symbol, labelled m13s, refers to a fit using the {\sc tlusty} (LTE) $+$ {\sc synspec} (LTE) grid ({\bf m13n10}), with $\xi_{\rm turb}=5$ \kmsec. 
        Formal errors are shown for the RSS fits. 
        Formal errors for the HRS fits are too small to show.
        } 
        \label{f:fits}
\end{figure}

\begin{figure}
\centering
 \includegraphics[width=0.48\textwidth,angle=180,trim={14cm 12cm 2.5cm 0cm},clip]{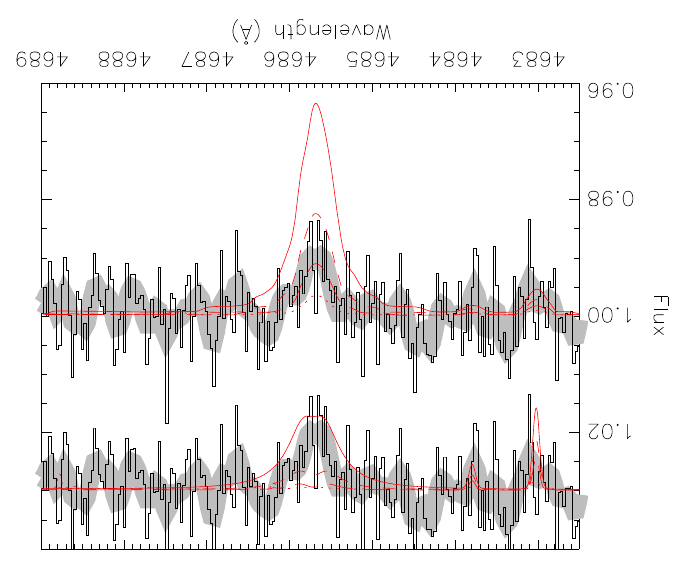}
        \caption{The HRS spectrum of \Zstar\ (histogram) in the vicinity of He{\sc ii}4686 compared with equivalent {\sc synspec} (top) and {\sc sterne} (bottom) models (red) having $T_{\rm eff}$ = 16 kK (dots), 18 kK (long dashes), 20 kK (short dashes) and 22 kK (solid). The thick grey curve shows the HRS spectrum smoothed with a 0.1\AA\ FWHM Gaussian filter. 
        } 
        \label{f:he2}
\end{figure}

\section[]{Stellar Atmosphere Analysis}
\label{s:fits}

\subsection{Model atmosphere grids.} In the process of fine analysis,  atmospheric parameters are obtained by finding a best-fit model spectrum, usually by interpolation in an appropriate grid. 
Typically, an estimate is made for the metal content and microturbulent velocity $\xi_{\rm turb}$, and then a grid of model atmospheres is computed to span a range of effective temperature $T_{\rm eff}$, surface gravity $g$ and hydrogen abundance $n_{\rm H}$ (fractional abundance by number). 
Since the composition and  $\xi_{\rm turb}$ are not known {\it ab initio}, the process is iterative. 
The first (SALT1) iteration used solar metallicity grids ({\bf p00}) 
\citep{jeffery21a}, from which $T_{\rm eff}\approx18\,500$K, $\log g/({\rm cm\,s^{-2}})\approx3.4$ and $n_{\rm H}\approx 0.01$ was deduced. Alternatively, $\log y \equiv \log n_{\rm He}/n_{\rm H} \approx 2$.
Predominantly weak metal lines suggested adoption of a metallicity scaled to 1/10 of the solar abundance ({\bf m10}). $\xi_{\rm turb}=0$\kmsec\ was assumed. 
{\sc sterne} models were calculated on grids defined by:\\
$T_{\rm eff} = 16, 18, 20, 22 $ kK,\\
$\log g/({\rm cm\,s^{-2}}) = 2.50, 2.75, 3.00, 3.25, 3.50, 3.75$, and \\
$n_{\rm H} = 0.00, 0.01, 0.03$.\\
For completeness, model grids were also calculated with metallicity scaled to 1/100 ({\bf m20}) and 1 times solar ({\bf p00}). 

The second iteration provided estimates of $T_{\rm eff}\approx20\,000$K, $\log g/({\rm cm\,s^{-2}})\approx3.5$ and $n_{\rm H}\approx 0.01$. 
From this, abundance estimates were obtained from measured equivalent widths, indicating an overall metallicity 1/20 times solar ($-1.3$\,dex = {\bf m13}), with the exception of nitrogen being close to $1\times$ solar ({\bf n10}).  
New models were calculated on the same grid as above using this custom mixture ({\bf m13n10}) and with $\xi_{\rm turb} = 0$, 5, 10 and 15\kmsec (abbreviated: v00, v05, v10 and v15).
Abundances obtained from this third iteration were within 0.3 dex of those used as inputs, so the process was deemed to have converged. 

LTE models were computed using {\sc tlusty} on the same grids and mixtures as the {\sc sterne} {\bf m10} and {\bf m13n10} grids. 
$\xi_{\rm turb} = 5$\kmsec{} was adopted throughout. 
Departure coefficients were computed for hydrogen and helium, and hence model spectra were obtained with hydrogen and helium out of LTE. All other atoms were treated in LTE throughout. 
The resulting grids are labelled  {\bf m10\_nlte} (or m10n) and {\bf m13n10\_nlte} (or m13n).

\begin{figure}
\centering

 \includegraphics[width=0.48\textwidth]{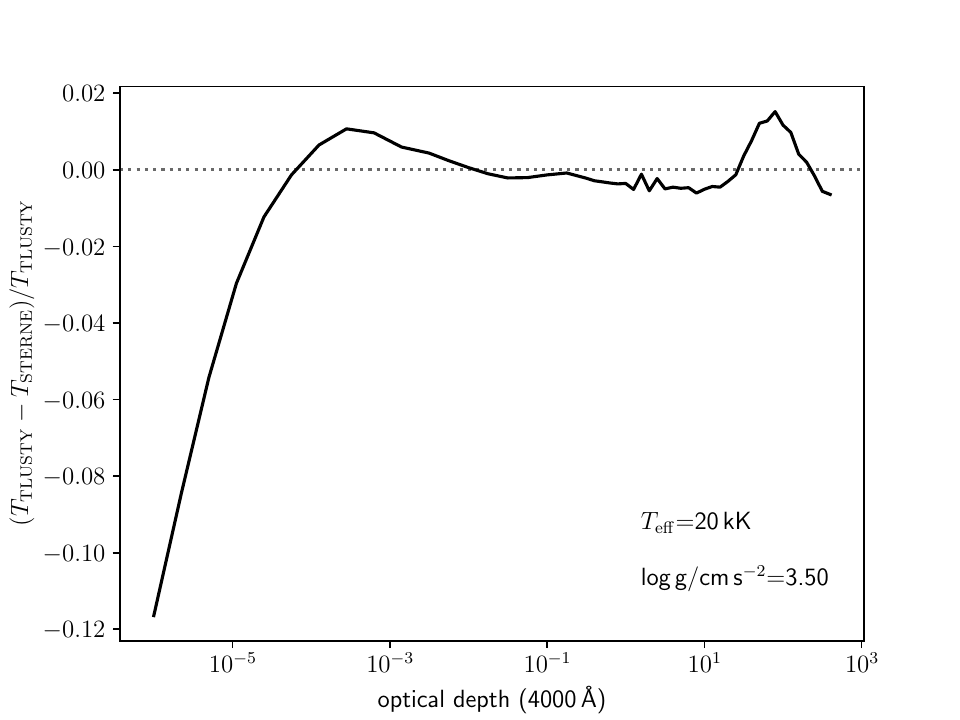}
        \caption{Comparison of temperature structures of {\sc tlusty} and {\sc sterne} model atmospheres in the sense $1 - (T_{\rm STERNE}/T_{\rm TLUSTY})$ as a function of optical depth.  
        Models are compared for $T_{\rm eff}=20\,000$\,K, $\log g/({\rm cm\,s^{-2}})=3.5$, $n_{\rm H}=0.01$, $\xi_{\rm turb}=5$\,\kmsec.  }
        \label{f:tdiff}
\end{figure}

\begin{figure*}
\centering
 \includegraphics[width=\textwidth,angle=0,trim={2.8cm 0cm 3.5cm 1.9cm},clip]{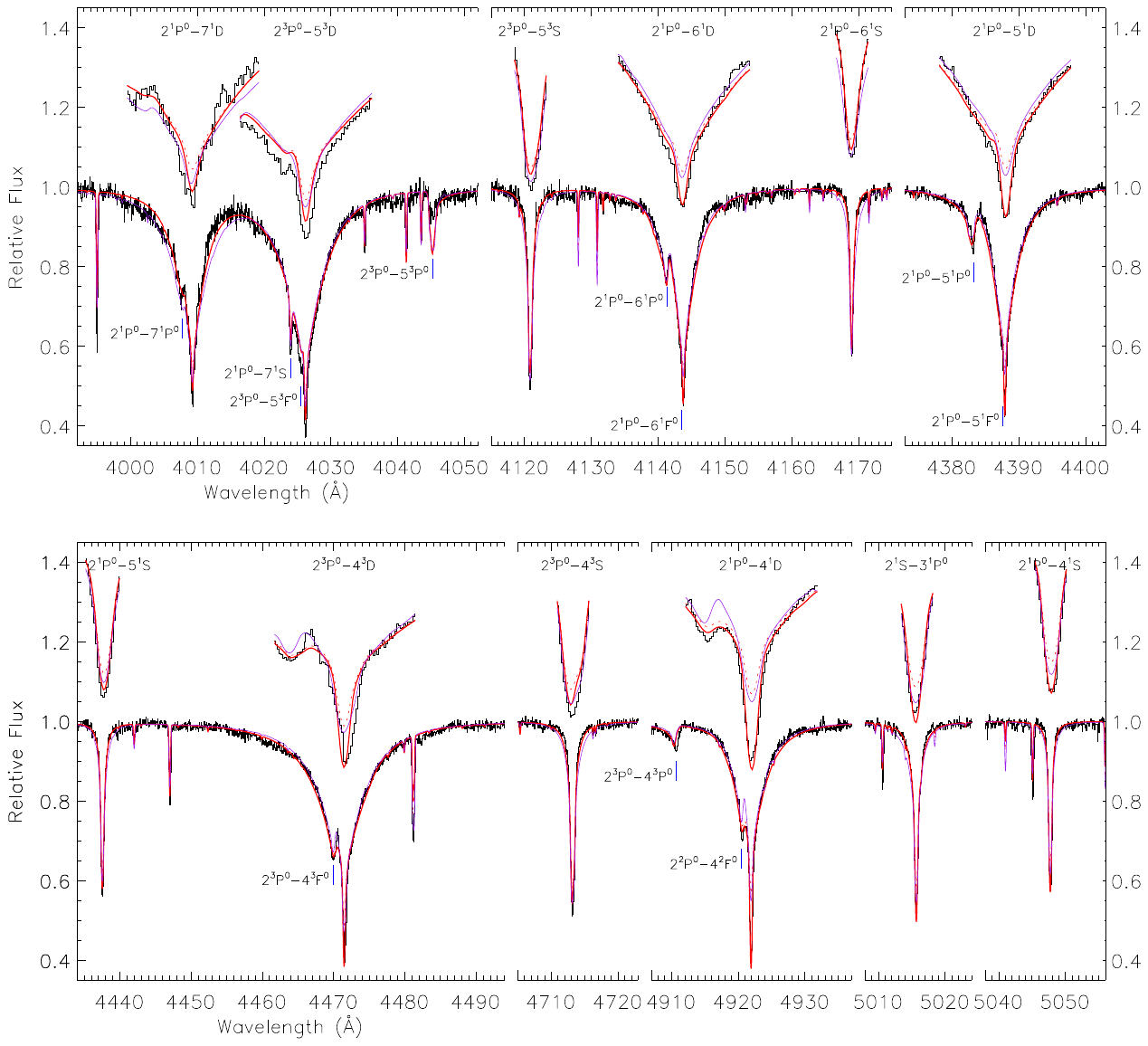}
        \caption{The HRS spectrum of \Zstar\ in the vicinity of several He{\sc i} lines compared with the best-fit non-LTE model (bold red) and the best-fit LTE models computed with {\sc tlusty+synspec} (dotted red) and {\sc sterne+spectrum} (solid violet). 
        The best-fit non-LTE model has $T_{\rm eff} = 20\,680 \pm 40$\,K,  $\log g /{\rm cm\,s^{-2}} = 3.49\pm0.01$,  and $n_{\rm H} = 0.0042\pm0.0001$ ($\log y = 2.37\pm0.01$) (formal errors).
        Each He{\sc i} line is labelled by its transition.
        The cores of strong lines are duplicated above, with the wavelength scale expanded by a factor 5. 
        Lines observed but not shown include $2^3P^0-4^3P^0$ 4517.4\AA, $2^1S-4^1P^0$ 3964.7\AA, $2^1P^0-8^1S$ 3935.9\AA, and $2^1P^0-8^1D$ 3926.5\AA.   
        } 
        \label{f:helium}
\end{figure*}

\subsection{Effective temperature and surface gravity.}

We use the $\chi^2$ minimizer {\sc sfit} \citep{jeffery01b} to optimise for  $T_{\rm eff}$, $\log g$ and $n_{\rm H}$ within a model grid, as defined by a given metal mixture and $\xi_{\rm turb}$. 
Other factors which affect the fit include the choice of wavelength range, bad data points, and local normalisation. 
The overall wavelength ranges used for the $\chi^2$ minimisation were  3900 -- 5070\AA\ (RSS) and 3900 -- 5200 \AA\ (HRS). 
Both spectra include many strong neutral helium and 4 hydrogen Balmer lines; the latter includes 2 additional strong He{\sc i} lines. 
The Ca{\sc ii} interstellar lines and other notably noisy regions of spectrum were excluded from the fit by setting very small weights to those regions. 
Key lines sensitive to temperature, gravity and hydrogen abundance were given extra weight in the HRS fits. 
These include H$\beta$, H$\gamma$, He{\sc i} 4471, 4388\AA, He{\sc ii} 4686\AA, Si{\sc ii} 4128, 4131\AA, and Si{\sc iii} 4553, 4568, 4575\AA. 

Solutions for $T_{\rm eff}$ and $\log g$ for each model grid and for  each spectrum are shown in Fig.\,\ref{f:fits}.  
The fits for $\log g$ are dominated by the pressure sensitive He{\sc i} lines and hence strongly correlate with $T_{\rm eff}$. $T_{\rm eff}$ is constrained by various line strengths and is sensitive to the model grid assumptions (metallicity, microturbulent velocity, etc). 

The top panel shows the position of the best fits to the RSS spectrum for 3 different model grids including a first generation {\sc sterne+spectrum} grid ({\bf m10}), a second generation  {\sc sterne+spectrum} grid ({\bf m13n10v05 $\equiv$ v05}), and a {\sc tlusty+synspec} grid ({\bf m13n}).
 
The light red triangles in the middle panel show positions of best fits to the HRS spectrum obtained with the same three grids ({\bf m10}, {\bf v05}, {\bf m13n}), and also positions of best fits with each of the other grids. 
These show the progression of fits as a function of metallicity (first generation: {\bf p00}, {\bf m10}, {\bf m20}), and microturbulent velocity (second generation: {\bf v00}, {\bf v05}, {\bf v10}, {\bf v15}). 
They thus demonstrate the magnitude of systematic errors arising from an incorrect metallicity or microturbulent velocity in the models. 
The bolder red triangles represent the three {\sc tlusty+synspec} grids ({\bf m10n}, {\bf m13n}, {\bf m13s}).

The bottom panel shows positions of the best fits to the HRS spectrum for each of the three reference grids ({\bf m10}, {\bf v05}, {\bf m13n}), but with the hydrogen abundance increased to 0.8\%.  

The general shape of the $\chi^2$ surface for each model grid is a narrow valley with a poorly-defined minimum. 
In addition to H and He{\sc i}, other lines affect precisely where the $\chi^2$ minimum occurs and hence fix \Teff. 
Examples include He{\sc ii} 4686\AA\ and Si{\sc ii} 4128, 4131\AA\, as well as the large number of N{\sc ii} lines.
The He{\sc ii} 4686\AA\ line in \Zstar\ is weak, with an equivalent width of  $4.8\pm2.4$\,m\AA. 
Nevertheless, its presence provides an important constraint on $T_{\rm eff}$.  
Fig.\,\ref{f:he2} compares the He{\sc ii} 4686\AA\ with models from the LTE
 {\bf m13n10} grid and from the non-LTE {\bf m10} grid covering $T_{\rm eff} = 16, 18, 20$ and 22 kK; $\xi_{\rm turb}=5$ \kmsec in both cases. 
 In the LTE grid, $T_{\rm eff} \approx 18$\,kK, whereas in the non-LTE grid, $T_{\rm eff} \approx 21$\,kK. These values are broadly consistent with the results in Fig.\,\ref{f:fits}.

\begin{figure*}
\centering
 \includegraphics[width=\textwidth,angle=0,trim={2.2cm 0.3cm 6cm 11cm},clip]{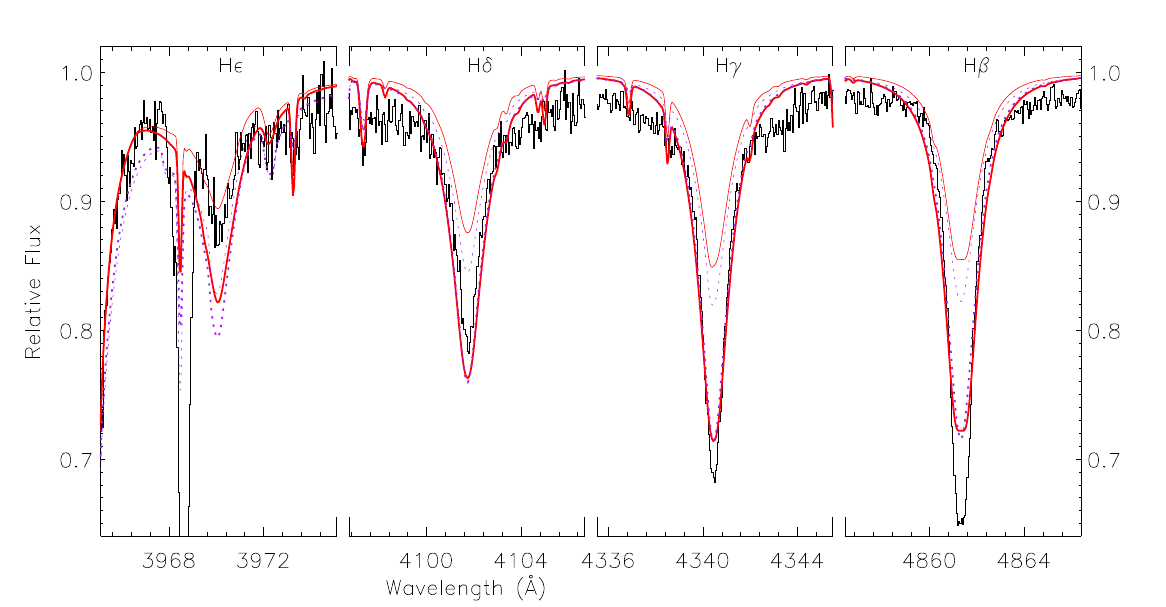}
        \caption{ The HRS spectrum of \Zstar\ in the vicinity of four Balmer lines compared with models for (a) a best-fit hydrogen abundance ($n_{\rm H}\approx0.004$; thin lines) and (b) a fixed value of the hydrogen abundance ($n_{\rm H}=0.008$; thick lines). Solid red lines show the best-fit with non-LTE hydrogen and helium ($\Teff=20.7$\,kK, $\sg=3.5$). Dotted violet lines show H and He in LTE ($\Teff=19.8$\,kK, $\sg=3.4$). The strong narrow line at 3968.5\AA\ is due to interstellar calcium. 
        } 
        \label{f:balmer}
\end{figure*}

\subsection{Departures from LTE}
\label{s:nlte}

A criticism of the LTE approximation is that it does not perform well in the cores of strong lines \citep{przybilla05,pandey11,kupfer17,pandey17}. 
This problem is mitigated by the use of models in which the hydrogen and helium level populations are treated in non-LTE (bold symbols in Fig.\,\ref{f:fits}).
For mixture {\bf m13n10}  and $\xi_{\rm turb}=5$\kmsec, the non-LTE solution (m13n) for the HRS spectrum is some 1400\,K hotter than the equivalent LTE solution (v05).  
It is also successful in fitting the He{\sc i} line cores much better than the LTE solution (Fig.\,\ref{f:helium}).  

Since the underlying models are computed with independent codes, it is necessary to verify that the difference comes primarily from the non-LTE treatment of  H and He lines, and not from other differences between the underlying models. 
Fig.\,\ref{f:tdiff} compares the temperature structures of models computed with {\sc sterne} and {\sc tlusty} for identical input parameters. 
Within the line-forming region ($10^{-4}<\tau_{4000}<1$) the temperatures differ by less than 1\%. 
A separate grid of models for the mixture mixture {\bf m13n10}  and $\xi_{\rm turb}=5$\kmsec was computed using {\sc tlusty} and {\sc synspec} in which the H and He lines were treated in LTE.  
Fitting the HRS spectrum to this grid  gave a solution almost identical to that for the equivalent grid calculated with {\sc sterne} and {\sc spectrum}.
This solution is labelled `m13s' in Fig.\,\ref{f:fits}. 
Equivalence between the {\sc tlusty+synspec} and {\sc sterne+spectrum} solutions in LTE is also demonstrated in Fig.\,\ref{f:helium}, particularly in the cores of the diffuse ($2P^0-nD$) lines. 

To address the question of how much the line cores affect the fits, the central 2\AA\ of all He{\sc i} lines were excluded.
For the reference {\sc sterne}$+${\sc spectrum} grid {\bf m13n10} (v05), $T_{\rm eff}$ and $g$ increased to almost exactly the values given by the {\sc tlusty} and {\sc synspec} (non-LTE) grid with the line cores included. 
However, excluding the line cores from the non-LTE grid increased $T_{\rm eff}$ by a similar amount $\approx 1\,000$\,K, implying that the line wings are also significantly different. 

\subsection{Formal and systematic errors}
It is difficult to assign confidences to our solutions; small changes in fit settings, such as wavelength range, wavelength weighting and normalization, can produce changes an order of magnitude larger than the formal errors in the $\chi^2$ minimization.
In most cases, we check for global minima by restarting the fit with different initial estimates and checking that same solutions are obtained. 
The formal errors for the HRS solutions have maxima of 50\,K in $T_{\rm eff}$ and 0.01\,dex in $g$. 
The correlation between $T_{\rm eff}$ and $g$ is robust over different spectra and different model grids. 

Microturbulent velocity is not too important for the $T_{\rm eff}, g$ solution so long as some line broadening is included; the results for  v05, v10 and v15 (mix {\bf m13n10}) are almost coincident, whilst being substantially hotter than for ${\bf v00}$. 
HRS results for {\bf m20}, {\bf m10} and {\bf p00} in Fig.\,\ref{f:fits} are diverse for decadal changes in metallicity but, if the metallicity is about right, other factors are probably more important. 
The LTE results for {\bf m10} and {\bf m13n10} (v05) are comparable, as are the non-LTE results for the same mixtures.

\subsection{RSS}
Solutions were obtained for the same model grids and for the RSS spectrum; for clarity only three solutions are shown in Fig.\,\ref{f:fits} (black diamonds), being for the {\bf m13n10} (diamond+square), {\bf m10} (diamond+cross) and {\bf m13n10\_nlte} (bold diamond) grids. 

In comparing HRS and RSS results, the influence of weak lines at lower resolution is reduced; exactly why this favours higher $T_{\rm eff}$ and $g$ is unclear.
Excluding Balmer lines and sharp He{\sc i} and H lines from the HRS fits {\it reduced} $T_{\rm eff}$ by 400 K, with no effect on gravity.

Both the LTE and  non-LTE RSS results are consistent with the  non-LTE HRS results.
The former seems counter-intuitive, but a test using the fully LTE {\sc tlusty} $+$ {\sc synspec} grid gave a much lower $T_{\rm eff}\approx18\,900$\,K, suggesting that systematic errors in fitting the RSS spectrum remain significant.

\subsection{Hydrogen abundance.} In a free solution, the hydrogen abundance measured from H$\beta, \gamma, \delta$ and $\epsilon$ is $\approx0.5\%$ by number. Forcing the solution to fit H$\gamma$ yields $\approx 0.8\%$. However, in both LTE and non-LTE solutions, the Balmer line fits are inconsistent in the sense that, if a model spectrum matches at H$\gamma$, the model for H$\beta$ is too weak and the models for H$\delta$ and H$\epsilon$ are increasingly too strong (Fig.\,\ref{f:balmer}).

The blue squares in Fig.\,\ref{f:fits} represent fits to the HRS spectrum in which $n_{\rm H}=0.008$ was held fixed. Otherwise the symbols match the corresponding red triangles; {\it i.e.} `$\times$' = {\bf m10}, `$\square$' = {\bf m13n10} v05, and a bold symbol = {\bf m13n10\_nlte}. 

 Both \citet{przybilla05} and \citet{pandey17} were able to resolve this `Balmer-line' problem through the use of non-LTE models. This has not been possible in the current case. 

\begin{figure}
\centering
 \includegraphics[width=0.49\textwidth,angle=180,trim={7.5cm 9.7cm 1.8cm 0cm},clip]{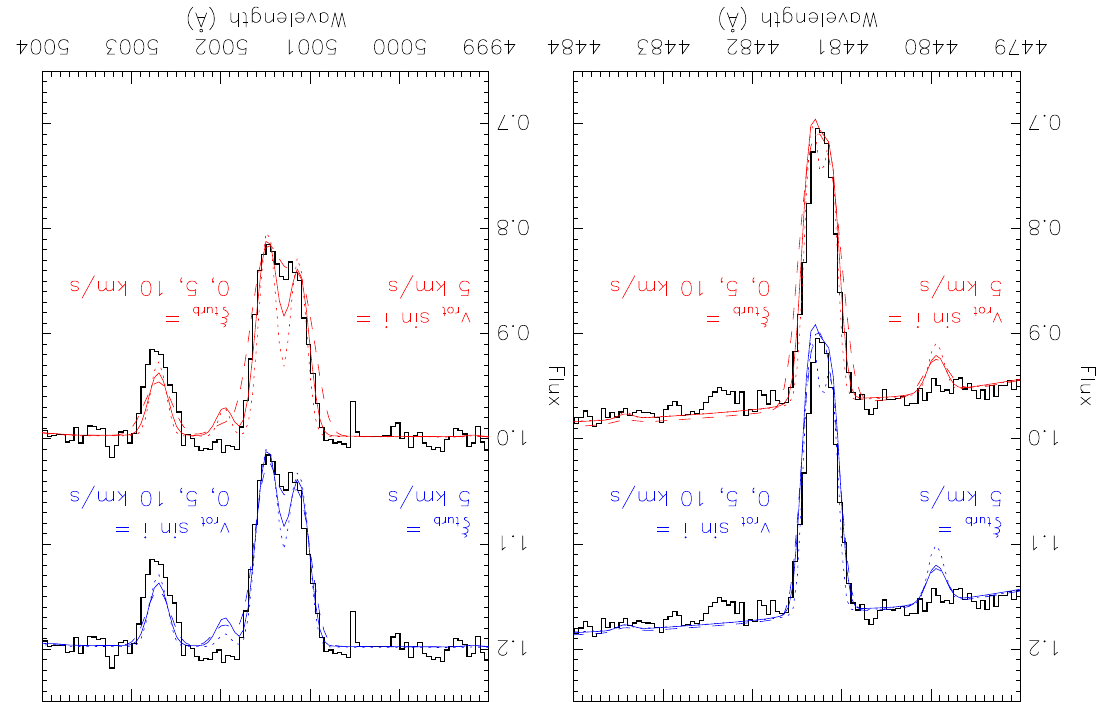}
        \caption{ The HRS spectrum of \Zstar\ in the vicinity of Mg{\sc ii}\,4481\AA\ and N{\sc ii}\,5001\AA\ doublets compared with a model spectrum for 3 different values of projected rotation velocity $v_{\rm rot} \sin i$  (top - blue) and microturbulent velocity $\xi_{\rm turb}$ (bottom - red). Values are indicated by dots $=0$\kmsec, solid lines $=5$\kmsec and  dashes $=10$\kmsec. 
        } 
        \label{f:vels}
\end{figure}

\begin{figure}
\centering
 \includegraphics[width=0.48\textwidth,angle=180,trim={11cm 12cm 2.5cm 0cm},clip]{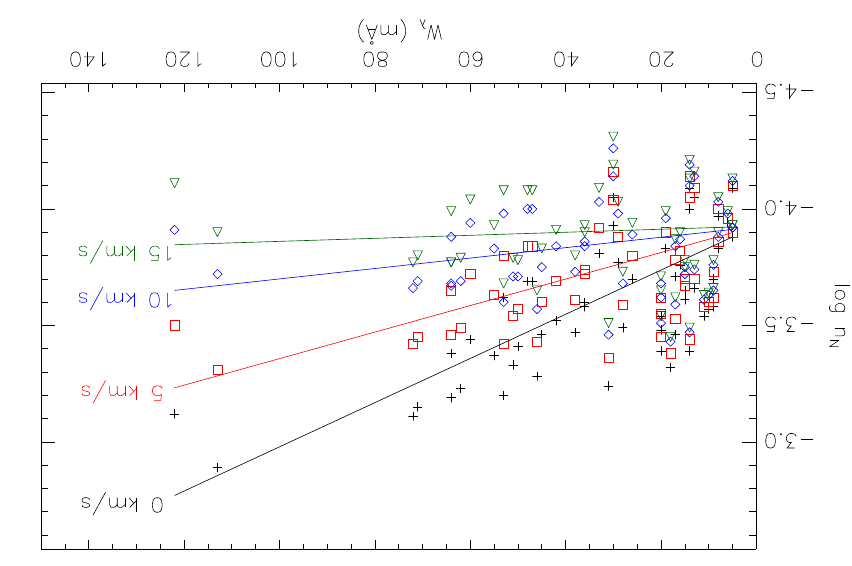}\\
 \includegraphics[width=0.47\textwidth]{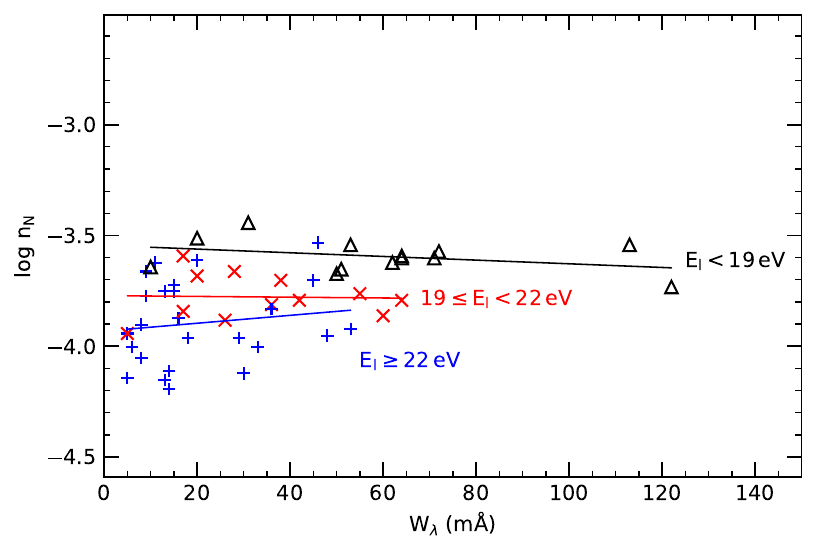}
        \caption{Top: Abundances derived from N{\sc ii} lines as a function of equivalent width for microturbulent velocities of $\xi_{\rm turb} = 0, 5, 10$ and 15\kmsec. The base model atmospheres used were {\bf m13n10/h01he99/t200g350} with input $\xi_{\rm turb} = 0$, 5, 10 and 15\kmsec. The solid lines show a linear regression for each value of $\xi_{\rm turb}$. 
        Bottom: The same for $\xi_{\rm turb} = 7$ \kmsec, with the N{\sc ii} lines separated into 3 groups ordered by $E_{\rm l}$ } 
        \label{f:n2}
\end{figure}

\subsection{Rotation.} The Mg{\sc ii} 4481\AA\ doublet is unresolved but asymmetric; a projected rotation velocity  $v \sin i = 5\pm2$\kmsec\ is sufficient to match the observed profile (Fig.\,\ref{f:vels}). 

\subsection{Microturbulent velocity.} 
\label{s:micro}
Two methods have been considered to estimate the microturbulent velocity $\xi$. 
 Looking at line profiles, the resolved N{\sc ii} 5001\,\AA\ doublet requires a value of $\xi_{\rm turb}$ between 5 and 10 \kmsec\  (Fig.\,\ref{f:vels}). 
The larger value produces profiles that are broader than observed; the lower value matches the line wings but over-resolves the doublet core. 
Alternatively, $\xi_{\rm turb}$ can be adjusted so that all of the N{\sc ii} lines yield a consistent nitrogen abundance across the full range of measured equivalent widths. 
This method indicates $\xi_{\rm turb} = 15\pm 5$\kmsec (Fig.\,\ref{f:n2}, upper panel).  
The conflict between these methods is partially resolved if the N{\sc ii} lines are considered according to the excitation potential of the lower energy level of the transition $E_l$. 
The lines fall naturally into three groups, namely lines with (1) $E_{\rm l}<19$eV, (2) $19{\rm eV}<E_{\rm l}<22$eV, and (3) $E_{\rm l}>22$eV. 
A value $\xi_{\rm turb}=7\pm1$\kmsec\ minimizes the absolute gradient ${\rm d} \log n_{\rm N}/{\rm d} W_{\lambda}$ averaged over all three groups (Fig.\,\ref{f:n2}: lower panel).   
However, the inconsistency between lines of different $E_l$ suggests that the temperature structure of the atmosphere is still not fully solved or that some lines may show non-LTE effects. 

Since the parameters derived from the spectral fits are relatively insensitive to $\xi_{\rm turb}>5$\kmsec, this value is deemed acceptable for the model atmosphere calculation. 
Its principal effect on the model structure is through line opacity in the ultraviolet which is reduced by the relatively low metallicity of the star. 

For measuring abundances, errors will not be too large if a low value of $\xi_{\rm turb}$ is adopted, so long as the line equivalent widths are small (e.g. $W_{\lambda}<50$m\AA).  
This would certainly be necessary if a spectral synthesis approach were to be used because otherwise the model profiles would be too broad to fit the observed profiles. 
On the basis of Fig.\,\ref{f:n2}, we have used $\xi_{\rm turb} = 7$\kmsec\ to measure abundances directly from measured equivalent widths using the individual line curve-of-growth method (Table B.2).
For reference, the majority of lines used in our analysis have equivalent widths $<100$m\AA. 
The minimum measurable equivalent width depends on the local signal-to-noise ratio which varies  throughout the \`echelle spectrum, but generally $W_{\lambda} > 4$m\AA.
This corresponds roughly to the line detection threshold $W_{\lambda}>n^2 \sigma_a^2 / \delta \lambda = 6.9$m\AA\ \citep{snowdon22} for $\sigma_{4540}=0.006$ (Table 1), pixel width $\delta \lambda = 0.05$\AA\ and a confidence level of 99\% ($n=3$).   

\subsection{Adopted solution.}

The final surface properties measured for \Zstar\ are  $T_{\rm eff} = 20\,700 \pm(40,250)$\,K  and  $\log g /{\rm cm\,s^{-2}} = 3.49\pm (0.01,0.03)$,  $n_{\rm H}=0.008\pm(0.0001,0.002)$  with $v \sin i = 5\pm2$\kmsec. 
 These are taken from the fit of the HRS spectrum to the {\sc tlusty} $+$ {\sc synspec} grid with mixture {\bf m13n10} and $\xi_{\rm turb}=5$\kmsec.
All elements are treated in LTE in the {\sc tlusty} models. H and He are treated in non-LTE in the {\sc synspec} formal solutions; other elements are in LTE.
1$\sigma$ errors are given in the form (statistical, systematic).  


\begin{table*}
\caption[Atmospheric abundances]
   {Atmospheric abundances of \Zstar, helium stars with similar $L/M$ ratios, and the Sun.
    Abundances are given as $\log \epsilon$, normalised to $\log \Sigma \mu \epsilon = 12.15$. A colon indicates an uncertain value. Ratios relative to the Sun are shown for \Zstar{} beneath. }
\label{t:abunds}
\small
\setlength{\tabcolsep}{3pt}
\begin{center}
\begin{tabular}{lrrrrrrrrrrrrrr}
Star     &    H &  He &  C &  N &  O & Ne & Mg & Al & Si &  P &  S & Ar &
Fe &  Ref \\[1mm]
 $\log \epsilon$   \\
{\Zstar}& 9.24 & 11.54 & 5.60 & 7.76 & 6.67 &  & 6.50 & 4.95 & 5.89 &  & 5.99 & & 6.11 &  \\
\multicolumn{1}{r}{$\pm$} & 0.10 &       & 0.22 & 0.19 & 0.12 &  & 0.12 & 0.15 & 0.12 &  & 0.25 & & 0.11 & \\[2mm]
{V652\,Her}     &9.61 &11.54&7.29 &8.69 &7.58&7.95 &7.80&6.12&7.47&:6.42&7.05& 6.64&7.04& 1,2  \\
{V652\,Her}$^a$ &9.5\,&11.5\,&7.0\,&8.7\,&7.6\,&8.1\,&7.1\,&  &7.4\,&   &7.4\,&    &7.1\,& 3  \\
{J1846$-$4138} & 9.56 & 11.54 & 6.91 & 8.69 & 7.78 & 8.29 & 7.91 & 6.44 & 7.61 & 5.62 & 6.93 &   & 7.07 & 4 \\
{BX\,Cir}       & 8.1\, &11.5\, &9.02 &8.4\,  &8.0\, &    &    &7.2\, &6.0\, &6.8\, &5.0\, &  6.6\,& 6.6\, & 5 \\
{BD\,$+10^{\circ}2179$} & 8.36 & 11.53 & 9.75 & 8.03 & 7.51 & 8.00 & 6.97 & 5.84 & 7.14 & & 6.84 & & 6.55 & 6 \\
{Sun}    &12\,\,\, &10.93 &8.43 &7.83 &8.69&7.93&7.60&6.45&7.51&5.45&7.12&6.40&7.50 & 7 \\[2mm]
 $\log \epsilon/\epsilon_{\odot}$    \\
{\Zstar} & $-2.76$ & 0.61 & $-2.83$ & $-0.07$ & $-2.02$ &  & $-1.10$ & $-1.50$ & $-1.62$ &  & $-1.13$ & & $-1.39$ & \\[2mm]
\end{tabular}
\end{center}
\parbox{170mm}{
References. 
1~\citet{jeffery99},
2:~\citet{jeffery01b},
3:~\citet{pandey17},
4:~\citet{jeffery17b},
5:~\cite{drilling98}, 
6:~\citet{kupfer17},
7:~\citet{asplund09}.
}
\parbox{170mm}{
Note. 
(a)~non-LTE measurements
}
\end{table*}

\subsection{Elemental Abundances}

Elemental abundances were computed for each measured absorption line using individual curves-of-growth computed with {\sc spectrum}. 
Two values were calculated for each line with model atmospheres from grid {\bf m13n10} having $\teff=20\,000$ and $22\,000$\,K, and otherwise $\log g=3.50$, $n_H=0.01$ and $\xi_{\rm turb}=5$\kmsec.  
Abundances were computed with $\xi_{\rm turb}=7$\kmsec, as indicated in \S\,\ref{s:micro}, and errors were propagated from the equivalent width errors to the abundances on a line-by-line basis. 
Line abundances and errors were interpolated to $\teff=20\,700$\,K.
Results for each line and each ion are shown in Appendix\,\ref{s:lines} (Table\,\ref{t:lines})\footnote{ 
Abundances are given in the form $\log \epsilon_i = \log n_i + c$ where $\log \Sigma_i a_i \epsilon_i = 12.15$,  $n_i$ are relative abundances by number with $\Sigma_i n_i = 1$,  $a_i$ are  atomic weights, and $c = \log \Sigma_i \epsilon_i$. 
This conserves values of $\epsilon_i$ for elements whose abundances do not change, even when the mean atomic mass of a mixture changes substantially, and conforms to the convention that $\log \epsilon_{\rm H} \equiv 12$ for the Sun and other hydrogen-normal stars.}.

The overall elemental abundances of measured species are given in Table\,\ref{t:abunds} and are compared with abundances for other relatively high-gravity extreme helium stars and the Sun. 
Where lines from more than one ion species are present, the overall elemental abundance is calculated from the weighted mean over all lines of that element. 

We note the very low abundances of magnesium, aluminium, silicon, sulphur and iron, indicative of a systemically low metallicity, $-1.53 < \log \epsilon_i/\epsilon_{\odot} < -0.87$, ${\rm Z}_i\in$(Mg,Al,Si,S,Fe). 
We have adopted $-1.3$ as an indicative mean. 

We also note the ratio C : N : O $\approx1:100:8$. 
This ratio is $4:90:7$ in V652\,Her, and $1:88:11$ in J1846$-$4138. 

Fluorine is an important stellar evolution tracer \citep{pandey06c,bhowmick20}. 
The crucial F{\sc ii} lines in the optical ultraviolet were undetectable in V652\,Her; 
\citet{bhowmick20} could only deduce an upper limit of 10 times the solar abundance. 
Our spectrum does not reach these lines. 

An independent check on the effective temperature is provided by the ionization equilibria of species where lines from more than one ion are present. 
Equilibrium is defined as the model effective temperature where the abundances derived from each ion would be equal. 
This was computed for Si\,{\sc ii/iii} yielding $T_{\rm ion} = 21\,550\pm 820$\,K  and for S\,{\sc ii/iii} yielding $T_{\rm ion} = 20\,000\pm 570$\,K.


\begin{figure}
    \includegraphics[width=0.49\textwidth]{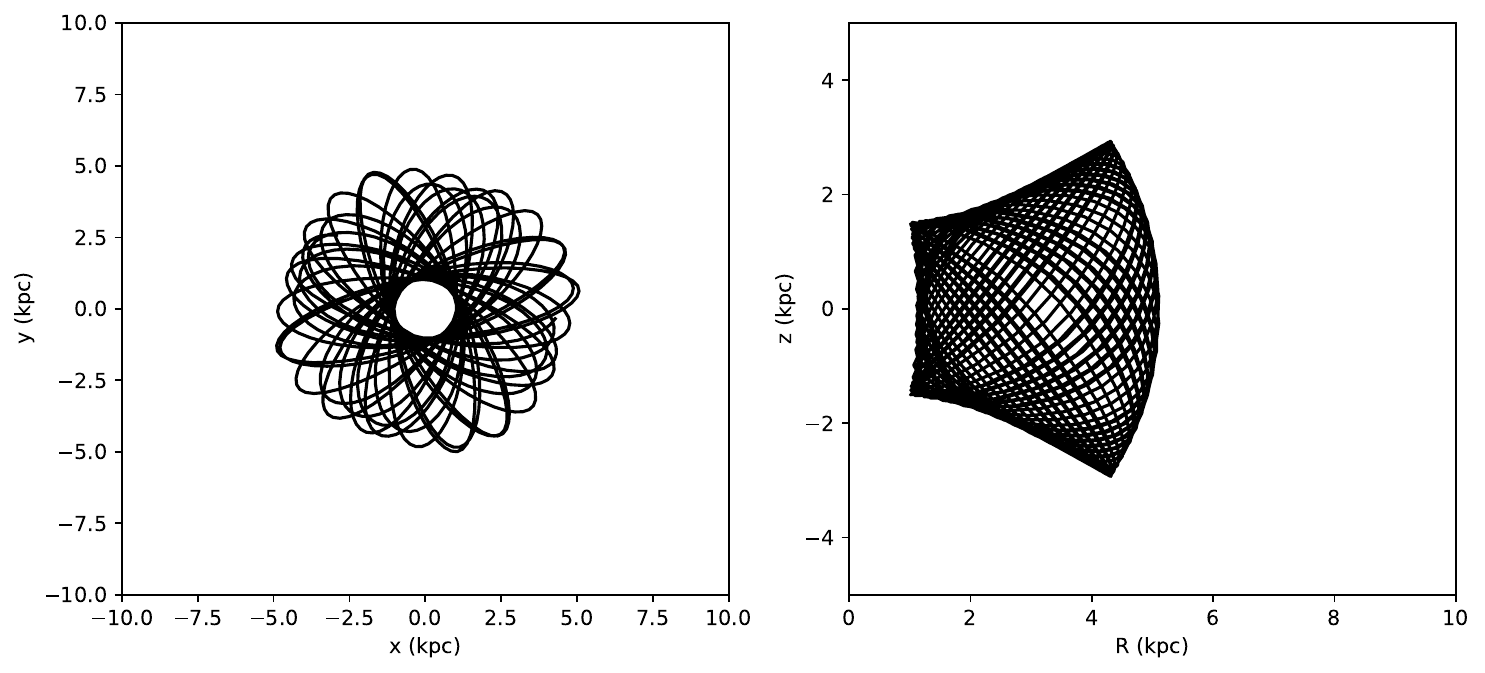}
    \caption{\texttt{galpy} orbit for \Zstar\ computed for 3 Gyrs from its current positions. 
    The panels show motion in the $\rm X-Y$ (left) and $\rm R-Z$ (right) planes, with the Galactic centre being at the origin. } 
    \label{f:galpy}
\end{figure}

\begin{figure}
 \includegraphics[width=0.48\textwidth]{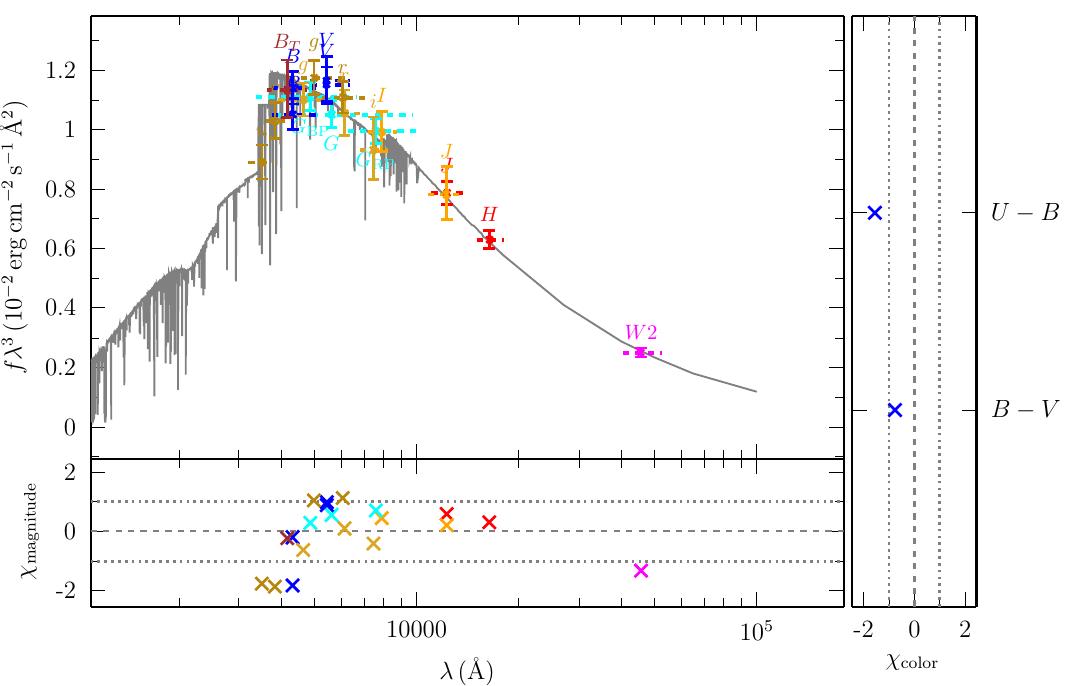}
        \caption{The spectral energy distribution of \Zstar\ and model fit (grey line) with $T_{\rm eff}=20\,700\,{\rm K}$ (upper panel). The lower panel shows the residuals.  Photometry is taken from: \, \citet{bianchi17}(FUV,NUV),\, \citet{hog00}(B$_{T}$,V$_{T}$),\, \citet{henden15}(B,V),\, \citet{riello21}($\rm G_{BP},\,G,\,G_{RP}$),\, \citet{onken19}(u,v,g,r,i,z),\, \citet{denis05}(I,J), \,\citet{cutri03}(J,H) and \citet{schlafly19}(W2).} 
        \label{f:sed}
\end{figure}

\section{Distance and Kinematics}

\subsection{Kinematics}

 From an analysis of EHe kinematics using {\it Gaia} eDR3 \citep{gaia16.mission,gaia21.dr3} data, \citet{pandey21} found that EHes belong to a spherical population that is more extended than the R\,CrB stars (RCBs) to which they were being compared.  
Extending earlier work by \citet{martin19.phd} and using positions and proper motions from {\it Gaia} DR3 \citep{gaia16.mission,gaia23.dr3}, \citet{philipmonai24} presented a more detailed study of EHe kinematics, distinguishing between the disk (thin/thick disk stars) and the  spherical components (halo/bulge stars). They found that EHes are present in all Galactic populations. 
\citet{tisserand24a} studied a large sample of RCBs, dustless hydrogen-deficient carbon stars (HDCs) and EHes. 
They came to no conclusion about the RCB distribution due to uncertainties arising from the {\it Gaia} point-spread function, but concurred that HdCs and RCBs are to be found in all four Galactic components. 
Both \citet{philipmonai24} and \citet{tisserand24a} support the idea that EHes formed recently from double white-dwarf mergers that arise originally from binary star formation over a range of epochs. 

\citet{philipmonai24} obtained a distance to \Zstar\  of  $4.4^{+1.2}_{-0.8}$\,kpc.  
Although the error is non-negligible ($\approx 22\%$), its retrograde orbit clearly makes \Zstar\ a halo member (Fig.\,\ref{f:galpy}).  
This is consistent with its low overall metallicity.

\subsection{Spectral Energy Distribution}

\citet{philipmonai24} used measurements of the spectral energy distribution (SED) to obtain radii and luminosities for all of the EHes.  
These measurements have been repeated for \Zstar\ using a more appropriate grid of model atmospheres ({\bf m13n10}) and \teff\ and \sg\ as given by the model atmosphere analysis (Fig.\,\ref{f:sed}).  
The results are only marginally different to those given by \citet{philipmonai24}.
With color excess $E(44-55)=0.067\pm 0.011$mag being the only free parameter in the SED fit, the angular diameter is given as $\log \theta / {\rm rad} = -10.494\pm0.007$. 
At the {\it Gaia} distance, this yields a median radius of $3.4^{+1.0}_{-0.6}$\Rsolar\ and luminosity $1.9^{+1.2}_{-0.6}\times10^3$\Lsolar.
With substantial errors in both distance and gravity, the median mass of $1.3^{+0.8}_{-0.5}$\Msolar is not well constrained. 
Using modes rather than medians reduces all of the above values by some 30\%. 

\citet{philipmonai24} concluded that EHes may be divided into high- and low-luminosity groups, which may be associated with different types of progenitor, namely CO+He and He+He double white dwarf mergers respectively. 
Both groups include members from the galactic thin disk, the galactic thick disk and the galactic halo.

\Zstar\ is marginal in terms of luminosity, being at the upper limit of the low-luminosity group.
However, its high nitrogen abundance strongly suggests that \Zstar\ belongs to the low luminosity group, which also includes the N-rich EHe stars V652\,Her and J1846$-$4138.
C-poor and N-rich surfaces are only expected for some He+He mergers, and not for He+CO mergers.
C-rich and N-rich surfaces are expected for He+CO mergers and the more massive He+He mergers \citep{zhang12a}.

\begin{figure*}
       \includegraphics[width=0.8\textwidth,trim={1cm 0cm 1cm 12cm},clip]{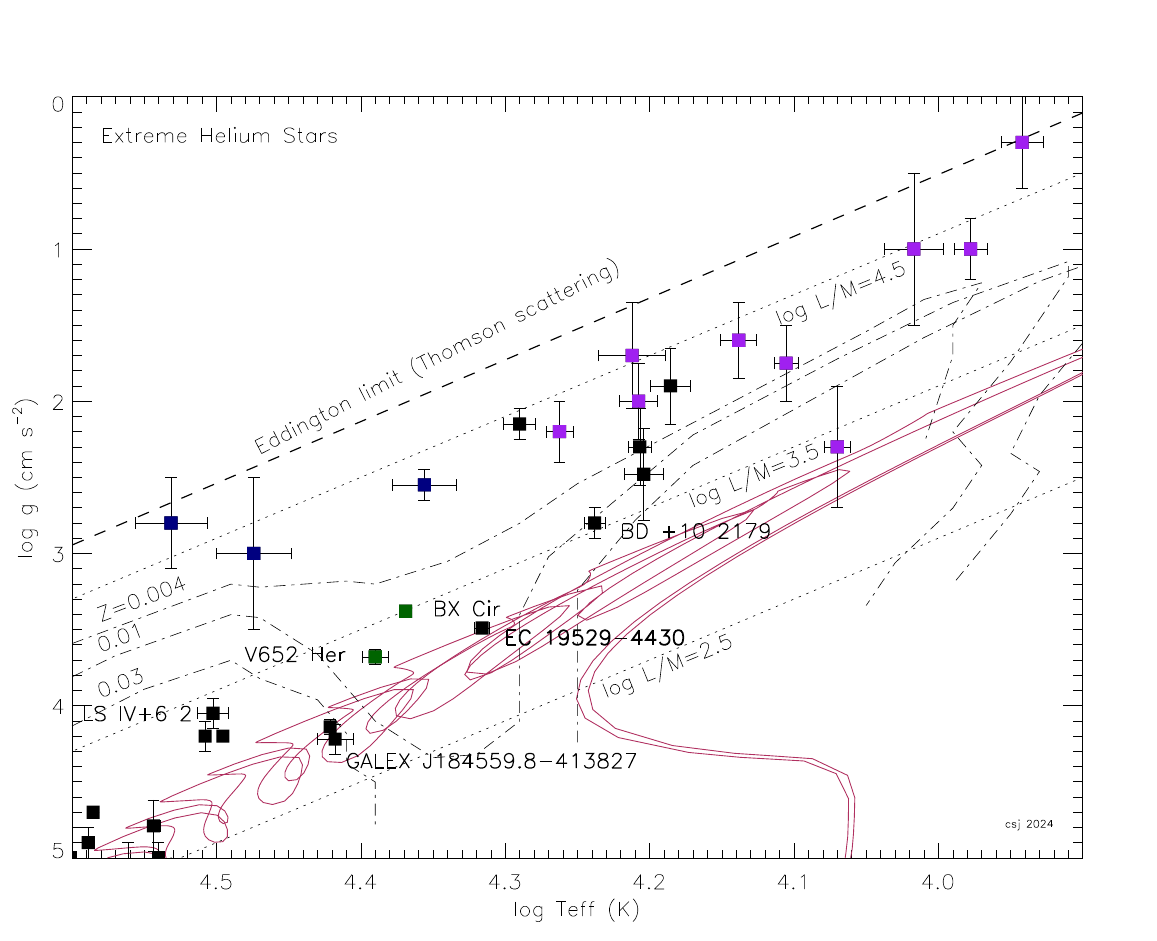}
        \caption{The $g-T_{\rm eff}$ diagram for \Zstar\ with reference to other known extreme helium stars. The positions of the Eddington limit (Thomson scattering: dashed), 
luminosity-to-mass contours (solar units: dotted) and lower boundaries for pulsation 
instability (metallicities $Z$\,=\,0.004, 0.01, 0.03: dot-dashed) \citep{jeffery99} are also shown.
Variable EHes are shown in purple (cool), blue (hot) and green (V652\,Her like variables). Non-variables are black. 
Data for $T_{\rm eff},g$ are as in \citet{jeffery08.ibvs}, \citep{ahmad03a} or \citet{jeffery21a}, except for BD+10$^{\circ}$2179 \citep{kupfer17},  \Jstar\ \citep{jeffery17b}, BPS\,CS\,22940--0009 \citep{snowdon22} and \Zstar. 
Post-merger evolution tracks for models of 
He+He white dwarf mergers \citep[0.30+0.25 and 0.30+0.30\,$M_\odot$]{zhang12a} are shown in red.
Stars from Table\,\ref{t:abunds} and other selected stars from the low-mass sequence identified by \citet{philipmonai24} are labelled. 
 } 
        \label{f:tracks}
\end{figure*}

\section{Conclusion}

We have presented a fine analysis of the spectrum of the  recently discovered extreme helium star \Zstar.
In doing so, we have examined the impact of several assumptions inherent in the model atmosphere approach, including the metallicity, the microturbulent velocity and departures from LTE. 
All three, if not correctly treated, can have a significant impact on the final result, amounting cumulatively in this case to over 3\,000\,K (or $>10\%$) in \teff. 
Treating the neutral helium lines in non-LTE, and metal line-blanketing with an appropriate metallicity and microturbulent velocity, we obtain a solution for \teff\ and \sg\ which is consistent over several discriminants. 
Citing the systematic errors, we conclude that $T_{\rm eff} = 20\,700 \pm250$\,K and $\log g /{\rm cm\,s^{-2}} = 3.49\pm0.03$. 
The surface hydrogen abundance cannot be obtained consistently from all four Balmer lines measured; 
0.8\% by number is given by the H$\gamma$ line, but a mean value between 0.45\% and 1\% is possible.  
 Otherwise, the surface composition is predominantly metal-poor, with $\log \epsilon_{\rm Fe}/\epsilon_{\odot}=-1.39$, or $\langle \log \epsilon_i/\epsilon_{\odot}\rangle \approx-1.3$ for  ${\rm Z}_i\in$(Mg,Al,Si,S,Fe). 
For such a metal-poor mixture, nitrogen is 1.2 dex overabundant.
Since carbon and oxygen are 1.5 and 0.7 dex underabundant, respectively, with respect to the mean metallicity, the surface appears to be composed primarily of CNO-processed helium.  
Kinematically, \Zstar\ is a galactic halo star in a retrograde orbit. 

Fig.\,\ref{f:tracks} shows \Zstar\ relative to other EHes in a $\log g - \log \teff$ diagram. 
Contours of similar $L/M$ ratio run diagonally across this figure.
\citet{philipmonai24} argued that the distribution of EHes in this figure suggests the presence of low and high $L/M$ sequences. 
The former are consistent with evolutionary tracks for merging He+He white dwarfs \citep{zhang12a}. 
These form originally from binary star systems in which both components expand at the end of core-helium burning.  
If sufficiently close, they will interact in a common envelope. 
One or more common-envelope phases will remove material from both stars and shrink the orbit to leave a double white dwarf (DWD) binary.  
The overall age of \Zstar\ is thus dominated by the main-sequence lifetime of the progenitor binary and the gravitational wave radiation timescale between DWD formation and merging. 
Both are determined by the mass ratios and separations of the initial main-sequence stars and of the two white dwarfs. 
On the basis of binary-star population synthesis calculations by \citet{yu10,yu11}, \citet{philipmonai24} argued that approximately 50\% of EHes formed in the current epoch should come from He+He WD binaries in the galactic halo. 
\citet{zhang12a} demonstrated that the more massive He+He merger products should be carbon-rich {\it and} nitrogen-rich, whilst the less massive products would {red carbon-poor and} nitrogen-rich.
This model therefore provides a good description of \Zstar, consistent with its surface composition, galactic orbit, and position in the $\log g - \log \teff$ diagram. 
 
\Zstar\ is the most metal-poor EHe star known. 
It is the coolest member of the carbon-poor group, of which 
V652\,Her and J1846-4138 are slightly warmer metal-rich analogues.
Hotter  carbon-poor EHe stars are found as subdwarfs.
The non-detection of pulsations is probably associated with the iron abundance being too low to drive pulsations. 
There is no evidence for a binary companion either from radial velocities or the spectra energy distribution. 
It is most likely that \Zstar\ formed from the merging of two helium white dwarfs, which themselves formed as a binary system some 11 Gyr ago, and that it will evolve to become a core helium-burning EHe subdwarf. 
It will be important to identify cooler N-rich EHe stars, additional metal-poor EHe stars, and to study other chemical species, such as neon, to better understand the He+He WD merger process.  

\section*{Acknowledgments}

The authors thank the referee for valuable comments and in particular for helping to resolve a discrepancy between two different measurements of the microturbulent velocity.  

The Armagh Observatory and Planetarium is funded by direct grant from the Northern Ireland Department for Communities.
CSJ and LJS acknowledge support from the UK Science and Technology Facilities Council (STFC) Grant No. ST/M000834/1. 

The spectroscopic observations reported in this paper were obtained using the Southern African Large Telescope (SALT)
under programs 2018-2-SCI-033, 2018-2-SCI-033 and 2021-1-MLT-005  (PI: Jeffery). 
The authors are indebted to the work of the entire SALT team.

This paper reports on data collected with the TESS mission, obtained from the MAST data archive at the Space Telescope Science Institute (STScI). Funding for the TESS mission is provided by the NASA Explorer Program. STScI is operated by the Association of Universities for Research in Astronomy, Inc., under NASA contract NAS 5-26555.

\section*{Data Availability}
The SALT and TESS data reported in this paper are available in their respective public archives ({\tt ssda.salt.ac.za}, and {\tt archive.stsci.edu/missions-and-data/tess}). Theoretical model atmospheres and spectra generated for this paper will be made available on reasonable request to the principal author. 

\bibliographystyle{mn2e}
\bibliography{ehe}

\appendix
\renewcommand\thefigure{A.\arabic{figure}} 
\renewcommand\thetable{A.\arabic{table}} 

\section[]{Model Atmosphere Grids}
 Details for each of the model atmosphere grids used in \S\,\ref{s:fits} are given in Table\,\ref{t:grids}.

\begin{table*}
\caption[Model Atmosphere Grids]
   {Identifiers, chemical composition and other parameters for model atmosphere grids used in \S\,4. Where different, labels used in Fig.\,\ref{f:fits} are given in parentheses.} 
\label{t:grids}
\small
\begin{center}
\begin{tabular}{llcrccl}
Label           &   Code    &  $n_{\rm N}$ & [Fe/H] &  $\xi_{\rm turb}$ & Lines & Comment \\
{\bf p00}       & {\sc sterne+spectrum} & $\odot$     & 0.0    & 5  &           &  LTE + scattering \\
{\bf m10}       & {\sc sterne+spectrum} & $\odot/10$  & $-1.0$ & 5  &           &  LTE + scattering \\
{\bf m20}       & {\sc sterne+spectrum} & $\odot/100$ & $-2.0$ & 5  &           &  LTE + scattering \\
{\bf m13n10v00} (v00) & {\sc sterne+spectrum} & $\odot$     & $-1.3$ & 0  &           &  LTE + scattering \\
{\bf m13n10v05} (v05)& {\sc sterne+spectrum} & $\odot$     & $-1.3$ & 5  &           &  LTE + scattering \\
{\bf m13n10v10} (v10)& {\sc sterne+spectrum} & $\odot$     & $-1.3$ & 10 &           &  LTE + scattering \\
{\bf m13n10v15} (v15)& {\sc sterne+spectrum} & $\odot$     & $-1.3$ & 15 &           &  LTE + scattering \\
{\bf m10\_nlte} (m10n) & {\sc tlusty+synspec}  & $\odot/10$  & $-1.0$ & 5  &           &  model LTE, H + He{\sc i} lines nLTE  \\
{\bf m13n10\_nlte} (v05n) & {\sc tlusty+synspec}  & $\odot$  & $-1.3$ & 5  &           &  model LTE, H + He{\sc i} lines nLTE  \\
{\bf m13n10\_lte} (v05s) & {\sc tlusty+synspec}  & $\odot$  & $-1.3$ & 5  &           &  LTE  \\
\end{tabular}
\end{center}
\end{table*}

\renewcommand\thefigure{B.\arabic{figure}} 
\renewcommand\thetable{B.\arabic{table}} 

\begin{table}
\caption{ Equivalent widths and scaled abundances for 
unblended absorption lines in the spectrum of \Zstar. 
} 
\label{t:lines}
\begin{flushleft}
\begin{tabular}{lrrrrrr}
\hline
Ion & \multicolumn{2}{r}{Reference} \\
$\lambda$ & $E_{\rm l}$ & $\log gf$ & $W_{\lambda}$ & $\sigma_{W_{\lambda}}$ & $\log \epsilon_i$ & $\sigma_{\log \epsilon_i}$  \\ 
\AA & eV  &    & m\AA & m\AA  & \\ 
\hline
\ion{N}{ii} & \multicolumn{5}{l}{\citet{becker89d}} \\
3919.01 & 20.410 & $-$0.335 & 36 & 20 & 7.70 & 0.38 \\
3955.85 & 18.466 & $-$0.780 & 50 & 20 & 7.85 & 0.31 \\
3995.00 & 18.498 & 0.225 & 122 & 26 & 7.78 & 0.29 \\
4035.08 & 23.120 & 0.597 & 30 & 12 & 7.38 & 0.25 \\
4041.31 & 23.128 & 0.830 & 53 & 21 & 7.57 & 0.31 \\
4056.90 & 23.140 & $-$0.461 & 11 & 8 & 7.89 & 0.36 \\
4073.05 & 23.124 & $-$0.160 & 20 & 11 & 7.89 & 0.32 \\
4082.27 & 23.132 & 0.150 & 19 & 11 & 7.55 & 0.34 \\
4171.60 & 23.196 & 0.280 & 14 & 7 & 7.31 & 0.27 \\
4173.57 & 23.240 & $-$0.457 & 5 & 4 & 7.57 & 0.38 \\
4176.16 & 23.196 & 0.600 & 33 & 10 & 7.49 & 0.21 \\
4179.67 & 23.250 & $-$0.204 & 13 & 8 & 7.76 & 0.30 \\
4195.97 & 23.242 & $-$0.290 & 8 & 6 & 7.61 & 0.36 \\
4199.98 & 23.246 & 0.030 & 16 & 7 & 7.63 & 0.24 \\
4227.74 & 21.600 & $-$0.089 & 26 & 12 & 7.62 & 0.29 \\
4236.93 & 23.239 & 0.383 & 46 & 13 & 7.96 & 0.22 \\
4237.05 & 23.242 & 0.553 & 45 & 13 & 7.79 & 0.22 \\
4241.18 & 23.240 & $-$0.336 & 9 & 5 & 7.73 & 0.29 \\
4241.76 & 23.242 & 0.210 & 47 & 12 & 7.54 & 0.20 \\
4242.44 & 23.240 & $-$0.336 & 6 & 4 & 7.51 & 0.32 \\
4417.10 & 23.415 & $-$0.340 & 9 & 5 & 7.84 & 0.29 \\
4427.24 & 23.420 & $-$0.004 & 8 & 9 & 7.45 & 0.54 \\
4431.82 & 23.410 & $-$0.152 & 5 & 4 & 7.36 & 0.39 \\
4432.74 & 23.410 & 0.595 & 29 & 10 & 7.53 & 0.22 \\
4433.48 & 23.420 & $-$0.028 & 15 & 5 & 7.77 & 0.19 \\
4442.02 & 23.420 & 0.324 & 14 & 5 & 7.39 & 0.21 \\
4447.03 & 20.411 & 0.238 & 64 & 10 & 7.70 & 0.14 \\
4530.40 & 23.470 & 0.671 & 36 & 6 & 7.65 & 0.13 \\
4601.48 & 18.468 & $-$0.385 & 71 & 14 & 7.90 & 0.19 \\
4607.16 & 18.464 & $-$0.483 & 64 & 15 & 7.90 & 0.21 \\
4613.87 & 18.468 & $-$0.607 & 51 & 11 & 7.85 & 0.17 \\
4621.29 & 18.468 & $-$0.483 & 62 & 10 & 7.88 & 0.13 \\
4630.54 & 18.484 & 0.093 & 113 & 10 & 7.96 & 0.12 \\
4643.09 & 18.484 & $-$0.385 & 72 & 10 & 7.93 & 0.13 \\
4667.21 & 18.497 & $-$1.580 & 10 & 4 & 7.87 & 0.18 \\
4678.14 & 23.572 & 0.434 & 13 & 4 & 7.34 & 0.16 \\
4694.70 & 23.570 & 0.111 & 15 & 5 & 7.74 & 0.19 \\
4774.24 & 20.650 & $-$1.055 & 5 & 2 & 7.57 & 0.25 \\
4779.72 & 20.650 & $-$0.577 & 20 & 6 & 7.82 & 0.17 \\
4788.13 & 20.650 & $-$0.388 & 28 & 8 & 7.83 & 0.18 \\
4803.29 & 20.660 & $-$0.135 & 38 & 8 & 7.79 & 0.14 \\
4987.38 & 20.940 & $-$0.630 & 17 & 4 & 7.90 & 0.15 \\
5001.13 & 20.646 & 0.282 & 55 & 8 & 7.72 & 0.23 \\
5001.47 & 20.654 & 0.452 & 60 & 9 & 7.62 & 0.13 \\
5002.70 & 18.480 & $-$1.086 & 31 & 6 & 8.06 & 0.12 \\
5007.33 & 20.940 & 0.161 & 42 & 7 & 7.69 & 0.13 \\
5010.62 & 18.470 & $-$0.611 & 53 & 12 & 7.95 & 0.18 \\
5025.66 & 20.666 & $-$0.438 & 17 & 5 & 7.65 & 0.17 \\
5045.09 & 18.460 & $-$0.389 & 65 & 14 & 7.90 & 0.19 \\
5073.59 & 18.497 & $-$1.280 & 20 & 7 & 7.99 & 0.20 \\
\cmidrule{6-7}
 &  &  &  &  & 7.76 & 0.19 \\
\cmidrule{6-7}
 &  &  &  &  & \multicolumn{2}{r}{$\mathrm{\Delta}\epsilon=-0.011$} \\
\hline
\end{tabular}
\end{flushleft}
\end{table}

\begin{table}
\addtocounter{table}{-1}
\caption{(cont.)}
\begin{flushleft}
\begin{tabular}{lrrrrrr}
\hline
Ion & \multicolumn{2}{r}{Reference} \\
$\lambda$ & $E_{\rm l}$ & $\log gf$ & $W_{\lambda}$ & $\sigma_{W_{\lambda}}$ & $\log \epsilon_i$ & $\sigma_{\log \epsilon_i}$  \\ 
\AA & eV  &    & m\AA & m\AA  & \\ 
\hline
\ion{C}{ii} &  \multicolumn{5}{l}{\citet{yan87}} \\
4267.02$\rceil$ & 18.047 & 0.559 &    &    &      & \\
4267.27$\rfloor$ & 18.047 & 0.734 & 18 & 8 & 5.60 & 0.22 \\
\cmidrule{6-7}
& &  &  &  & 5.60 & 0.22 \\
\cmidrule{6-7}
& &  &  &  & \multicolumn{2}{r}{$\mathrm{\Delta}\epsilon=0.005$}\\

\ion{O}{ii} &  \multicolumn{5}{l}{\citet{bell94}} \\
3973.26 & 23.442 & $-$0.015 & 8 & 11 & 6.81 & 1.07 \\
4072.15 & 25.650 & 0.552 & 5 & 5 & 6.58 & 0.90 \\
\cmidrule{6-7}
& &  &  & & 6.67 & 0.12 \\
\cmidrule{6-7}
& &  &  & & \multicolumn{2}{r}{$\mathrm{\Delta}\epsilon=-0.016$}\\

\ion{Mg}{ii} & \multicolumn{5}{l}{\citet{wiese69}} \\
4481.13$\rceil$ & 8.863 & 0.568 &  &  &  &  \\
4481.15\,$\arrowvert$ & 8.864 & $-$0.560 &   &  &  &  \\
4481.33$\rfloor$ & 8.863 & 0.732 & 110 & 15 & 6.50 & 0.12 \\
\cmidrule{6-7}
& &  &  & & 6.50 & 0.12\\
\cmidrule{6-7}
& &  &  & & \multicolumn{2}{r}{$\mathrm{\Delta}\epsilon=0.035$}\\

\ion{Al}{iii} & \multicolumn{5}{l}{\citet{cunto93}} \\
4512.54 & 17.808 & 0.405 & 12 & 5 & 5.14 & 0.21 \\
4529.20 & 17.740 & 0.660 & 12 & 4 & 4.84 & 0.16 \\
\cmidrule{6-7}
& &  &  & & 4.95 & 0.15 \\
\cmidrule{6-7}
& &  &  & & \multicolumn{2}{r}{$\mathrm{\Delta}\epsilon=-0.004$}\\

\ion{Si}{ii} &  \multicolumn{5}{l}{\citet{becker90a}}\\
4128.07 & 9.837 & 0.369 & 18 & 9 & 6.02 & 0.28 \\
4130.87 & 9.839 & $-$0.824 & 22 & 9 & 5.93 & 0.21 \\
5041.03 & 10.070 & 0.262 & 6 & 3 & 5.73 & 0.26 \\
5055.98 & 10.070 & 0.517 & 8 & 3 & 5.60 & 0.22 \\
\cmidrule{6-7}
& &  &  & & 5.81 & 0.19 \\
\cmidrule{6-7}
& &  &  & & \multicolumn{2}{r}{$\mathrm{\Delta}\epsilon=0.031$}\\

\ion{Si}{iii} &  \multicolumn{5}{l}{\citet{becker90a}} \\
4552.62 & 19.018 & 0.283 & 77 & 12 & 6.36 & 0.14 \\
4567.82 & 19.018 & 0.061 & 55 & 10 & 6.31 & 0.14 \\
4574.76 & 19.018 & $-$0.416 & 30 & 7 & 6.34 & 0.14 \\
4716.65 & 25.330 & 0.491 & 3 & 2 & 5.96 & 0.34 \\
4813.30 & 25.980 & 0.702 & 4 & 2 & 6.05 & 0.28 \\
4819.72 & 25.980 & 0.814 & 5 & 2 & 5.99 & 0.26 \\
\cmidrule{6-7}
& &  &  & & 6.27 & 0.15 \\
\cmidrule{6-7}
& &  &  & & \multicolumn{2}{r}{$\mathrm{\Delta}\epsilon=-0.008$}\\

\ion{S}{ii} &  \multicolumn{5}{l}{\citet{wiese69}} \\
4815.55 & 13.672 & 0.177 & 7 & 3 & 5.87 & 0.17 \\
5032.41 & 13.668 & 0.176 & 9 & 4 & 6.01 & 0.19 \\
\cmidrule{6-7}
& &  &  & & 5.93 & 0.07 \\
\cmidrule{6-7}
& &  &  & & \multicolumn{2}{r}{$\mathrm{\Delta}\epsilon=0.037$}\\

\ion{S}{iii} &  \multicolumn{5}{l}{\citet{hardorp70}} \\
4253.59 & 18.244 & 0.400 & 17 & 7 & 5.80 & 0.26 \\
4284.98 & 18.193 & 0.110 & 12 & 6 & 5.87 & 0.27 \\
4332.69 & 18.188 & $-$0.240 & 4 & 4 & 5.67 & 0.46 \\
\cmidrule{6-7}
& &  &  & & 5.81 & 0.08 \\
\cmidrule{6-7}
& &  &  & & \multicolumn{2}{r}{$\mathrm{\Delta}\epsilon=-0.011$}\\

\ion{Fe}{iii} & \multicolumn{5}{l}{\citet{kurucz75}} \\
4164.73 & 20.634 & 0.972 & 5 & 4 & 5.95 & 0.38 \\
4419.60 & 8.241 & $-$2.207 & 10 & 5 & 6.18 & 0.24 \\
\cmidrule{6-7}
& &  &  & & 6.11 & 0.11 \\
\cmidrule{6-7}
& &  &  & & \multicolumn{2}{r}{$\mathrm{\Delta}\epsilon=0.011$}\\
\hline
\end{tabular}
\end{flushleft}
\end{table}

\section[]{Line Abundance Measurements}
\label{s:lines}

Table \ref{t:lines} shows the abundances obtained from individual absorption lines in the spectrum of \Zstar. 
{ Columns 1 and 4 give wavelengths and equivalent widths ($W_{\lambda}$)  for identified absorption lines. 
Multiplets treated as blends are grouped by a square bracket "]" following the wavelength. 
Columns 2 and 3 give the excitation potential of the lower level ($E_{\rm l}$) and the oscillator strength ($\log gf$) for each transition.
Column 5 gives the  $1\sigma$ error in $W_{\lambda}$, as described by \citet{snowdon22}. 
Lines with abundances $>2\sigma$  from the mean were rejected and omitted, and the means recalculated. 
The sources of oscillator strengths ($gf$) are given with the ion designation.
Column 7 gives the abundance $\log \epsilon_i$ deduced from each line.
Column 8 gives the error in the line abundance obtained by propagating $\sigma_{W_{\lambda}}$ through the abundance derivation in {\sc spectrum}.  
The mean abundance for each ion is given as the weighted mean of all lines shown.  
Errors on the mean abundance per ion are given as the weighted standard deviation where the number of lines $n>2$, the semi-range where $n=2$, or the line measurement error where $n=1$. 
The change in abundance due to a systematic error $\delta \teff = +250$K is shown beneath the mean abundance in the form $\mathrm{\Delta}\epsilon = {\rm d}\log \epsilon_i / {\rm d}T_{\rm eff} \times 250 {\rm K}$.

Abundances were calculated at the adopted temperature of $T_{\rm eff}=20\,700$\,K by interpolation, between LTE models with $\log g /{\rm cm\,s^{-2}} = 3.50$, $\xi_{\rm turb}=7$\,\kmsec{}, and $T_{\rm eff}$ of 20\,000 and 22\,000\,K. 

The N{\sc ii} lines fall into three natural groups by $E_{\rm l}$ (see Fig.\,\ref{f:n2}) with each group yielding a slightly different mean abundance. 
We compared the weighted mean from all lines taken together with the weighted mean for each group, namely lines with (1) $E_{\rm l}<19$eV, (2) $19{\rm eV}<E_{\rm l}<22$eV, and (3) $E_{\rm l}>22$eV and also groups (1)+(2) combined. 
Relative to the overall mean, the mean abundance for each group was (1) +0.21, (2) +0.01, (3) -0.10 and (1)+(2) +0.11 dex.
Although it contains the largest number of lines, group (3) is dominated by weak lines with large abundance errors; it therefore has a smaller impact on the overall mean.
Rather than reject any group of N{\sc ii} lines, we have adopted the overall  mean for the nitrogen abundance  

Abundances for sulphur and silicon are available for 2 ions in each case. 
The abundances for S{\sc ii} and S{\sc iii} differ by 0.12 dex, which is within the scatter of individual line measurements and less than the error in any individual line abundance. 
The abundances for Si{\sc ii} and Si{\sc iii} differ by 0.46 dex; the individual line abundance errors are less than this whilst the scatter for each ion is $\approx0.4$ dex. 
On this basis, Table\,\ref{t:abunds} gives the overall sulphur and silicon abundances from the weighted mean of all lines of both species . 
Since the silicon abundance is dominated by the strong Si{\sc iii} lines at 4553, 4568, and 4575\AA, which frequently do show strong non-LTE effects, we note that the silicon abundance may be overestimated.   
}

\label{lastpage}
\end{document}